\documentclass[amsmath,amssymb,aps,pra,reprint]{revtex4-1}
\usepackage{bbold}
\usepackage{upgreek}
\usepackage{dsfont}
\usepackage{graphicx}
\usepackage{braket} 
\usepackage{colortbl}

\newcommand{\bls}[1]{\mathbf{#1}} 

\newcommand{\rI}{\mathrm{I}}
\newcommand{\rR}{\mathrm{R}}

\newcommand{\ac}{\alpha_{\rm c}}

\begin{document}

\title{
	Gas-induced friction and diffusion of rigid rotors }

\author{Lukas Martinetz}
\author{Klaus Hornberger}
\author{Benjamin A. Stickler}
\affiliation{
 University of Duisburg-Essen, Faculty of Physics, Lotharstra\ss e 1, 47048 Duisburg, Germany}

\begin{abstract}
We derive the Boltzmann equation for the rotranslational  dynamics of an arbitrary convex rigid body in a rarefied gas. It yields as a limiting case the Fokker-Planck equation accounting for friction, diffusion, and  nonconservative drift forces and torques.
We provide the rotranslational friction and diffusion tensors for specular and diffuse reflection off particles with
spherical, cylindrical, and cuboidal shape,
and show that the theory describes thermalization, photophoresis, and the inverse Magnus effect in the free molecular regime.
\end{abstract}

\maketitle

\section{Introduction}

A particle moving and revolving in a rarefied gas experiences a nonconservative force and torque \cite{epstein1924} due to its random collisions with the surrounding gas atoms \cite{cercignani1988}. The resulting dynamics proved relevant for phenomena as diverse as the size distribution of dust grains in protoplanetary disks \cite{krauss2007,husmann2016}, the motion of satellites in the outermost layer of the atmosphere \cite{cook1964,prieto2014,curtis2013}, or the drag on dust particles in dirty plasmas \cite{shukla2015}. However, only recent experiments \cite{li2010,millen2014,kuhn2017a} in the field of levitated optomechanics \cite{chang2010,romeroisart2010} are capable of resolving the stochastic effect of individual scattering events.

An optically levitated nano- to microscale particle in high vacuum can be efficiently isolated from environmental disturbances, rendering it attractive for highly accurate measurements of force and torque \cite{ranjit2016,kuhn2017b} as well as for the observation of single-particle equilibration in a controlled environment \cite{li2010,millen2014,kuhn2017a,gieseler2014,hoang2017}. Cooling  the levitated object into the quantum regime \cite{li2011,gieseler2012,kiesel2013, asenbaum2013,millen2015,fonseca2016,vovrosh2017} will further increase the degree of accuracy and eventually allow tests of the quantum superposition principle for massive objects \cite{chang2010,romeroisart2010,romeroisart2011a,romeroisart2011b,bateman2014}.
Even before such interference tests will become available, the observed absence of collapse-induced heating \cite{nimmrichter2014,bahrami2014,laloe2014,diosi2015,goldwater2016,li2016,schrinski2017} can be used to falsify objective collapse models \cite{bassi2013,carlesso2016,helou2017}. 
All such experiments require a detailed understanding of how a nanoparticle is affected by the inevitable interaction with background gases.

In this article we provide a microscopic and comprehensive classical theory of the coupled translational and rotational dynamics of an arbitrarily shaped rigid rotor in a rarefied homogeneous background gas. We derive the rotranslational Boltzmann and Fokker-Planck equations in terms of the momentum transfer 
characterizing the scattering of single gas atoms at the individual surface points. This extends previous work for special interaction types and particle symmetries \cite{galkin2008}. Our results are directly applicable to recent experiments with  nonspherical levitated objects in high vacuum \cite{kane2010,kuhn2015,hoang2016,coppock2016,delord2017a,kuhn2017a,kuhn2017b}. Beyond optomechanics, the here derived equations may be relevant for explaining planet formation in interstellar dust \cite{krauss2007,husmann2016}, the interplanetary trajectory of elongated asteroids \cite{meech2017,fraser2018tumbling}, and dusty plasma dynamics  \cite{RevModPhys.81.1353}.

This paper is structured as follows: In Sec.~\ref{sec:be} we provide a microscopic derivation of the rotranslational   Boltzmann equation describing the dynamics of a particle in the free molecular regime for arbitrary shape and momentum transfer. In Sec.~\ref{sec:fpe} it is shown how the Fokker-Planck equation arises if the particle is slow in comparison to the surrounding gas atoms. We determine the nonconservative force and torque together with the resulting friction and diffusion tensors in Sec.~\ref{sec:snd} for the special cases of specular and diffuse reflection. The expressions are evaluated for spheres, cylinders, and cuboids in Sec.~\ref{sec:scc}, and we conclude in Sec.~\ref{sec:conc}.

\begin{figure*}
\centering
\includegraphics[width = 0.99\textwidth]{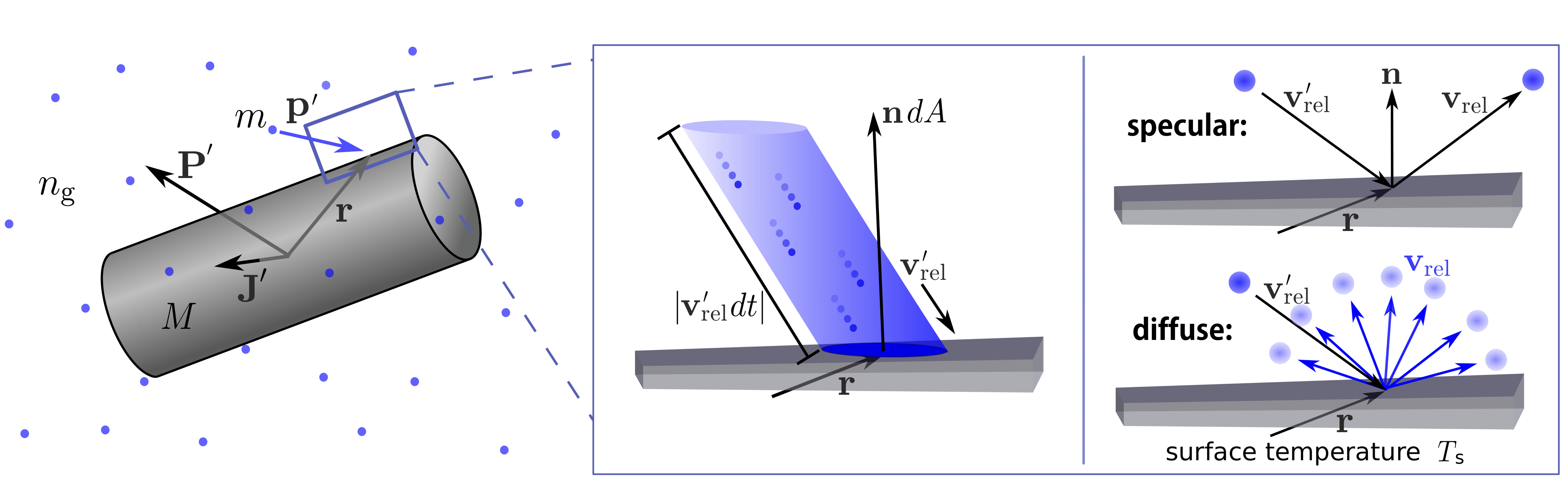}
\caption{Convex nanoparticle (gray) of mass $M$ with momentum ${\bf P}'$ and angular momentum ${\bf J}'$ moving through a dilute gas (blue) of density $n_{\rm g}$ and atomic mass $m$. The rate of scattering events off the infinitesimal surface element $dA$ at ${\bf r}$ depends on the incident relative velocity ${\bf v}_{\rm rel}'$ and the surface normal vector ${\bf n}$. The type of scattering process, such as specular and diffuse reflection, is incorporated by specifying a momentum transfer function.} \label{hauptbild}
\end{figure*}

\section{Rotranslational Boltzmann equation} \label{sec:be}

We consider an arbitrary convex rigid body of mass $M$ with moments of inertia $(I_1,I_2,I_3)$ moving through a homogeneous rarefied gas of number density $n_{\rm g}$ and atomic mass $m$. Denoting the rigid body degrees of freedom by $({\bf R},\Omega)$, with the center-of-mass position ${\bf R}$ and the orientation $\Omega$ (given e.g.\ by the Euler angles), one can write its tensor of inertia as $\rI(\Omega) = \rR(\Omega) \rI_0 \rR^{\rm T}(\Omega)$, where $\rR(\Omega)$ rotates from the initial orientation to $\Omega$ and $\rI_0 = {\rm diag}(I_1,I_2,I_3)$.

It is our aim to derive the dynamical equation of the rotranslational state, represented by the probability density $f_t(\bls{R},\Omega,\bls{P},\bls{J})$ with ${\bf P}$ and ${\bf J}$ the center-of-mass and angular momentum, respectively. In absence of collisions the conservative dynamics of the particle $\partial_t^{\rm cons} f_t$ follow from the Hamilton function $H = {\bf P}^2 / 2 M + {\bf J} \cdot \rI^{-1}(\Omega) {\bf J}/2 + V({\bf R}, \Omega)$ with the external potential $V({\bf R},\Omega)$. Whereas the conservative dynamics can be conveniently expressed in the phase space of canonically conjugate variables using the Poisson bracket, we will see that the gas-induced motion can be formulated much more efficiently in terms of the angular momentum vector ${\bf J}$. A detailed discussion of the transformation between $f_t$ and the canonical phase space distribution can be found in App.~\ref{app:tdf}.

Denoting the nonconservative change of state due to the surrounding gas by $\partial_t^{\rm coll}f_t$ the total time evolution is given by the rotranslational Boltzmann equation
\begin{equation}\label{eq:BE}
 \partial_t f_t = \partial^{\rm cons}_t f_t + \partial^{\rm coll}_t f_t.
\end{equation}
In what follows we will determine the rotranslational  Boltzmann collision integral $\partial^{\rm coll}_t f_t$ by considering individual scattering processes off the particle surface.

We will see that the essential ingredients are the particle shape, the gas distribution $\mu({\bf p})$ and the scattering dynamics of a single atom with the nanoparticle, expressed in terms of the distribution function $g({\bf q}|{\bf q}')$ for the reflected relative momentum ${\bf q}$ given the incident momentum ${\bf q}'$ and the position of impact.
(Sect.~\ref{sec:snd} provides $g({\bf q}|{\bf q}')$ for the special cases of specular and diffuse reflection.)

\subsection{Scattering rate and transfer function}

We assume the mean free path of the gas atom to exceed the extension of the particle. This is referred to as the {\it free molecular regime} or {\it Knudsen regime} \cite{epstein1924,cercignani1988}, and it is typically obeyed in optomechanical experiments and for objects in the outermost layer of the atmosphere. For particles with a convex surface we can thus take  the  impinging gas atoms to be independent of the motion of the particle and to be characterized only by their momentum distribution $\mu({\bf p})$. 

To define the collision rate consider the relative velocity between a gas atom with momentum ${\bf p}'$ and an infinitesimal surface element  $d{\bf A}$ at position ${\bf r}$,
\begin{equation}
\bls{v}_{\rm rel}'= \frac{\bls{p}'}{m} -\frac{\bls{P}'}{M} + \bls{r}\times\rI^{-1}(\Omega)\,\bls{J}',
\end{equation}
depending on the current nanoparticle orientation $\Omega$, its center-of-mass momentum ${\bf P}'$, and angular momentum ${\bf J}'$; see Fig.~\ref{hauptbild}. The number of atoms impinging the surface area  $dA$ per time increment $dt$ gives the infinitesimal collision rate,
\begin{equation}
d\Gamma=-n_{\rm g}d{\bf A}\cdot\bls{v}_{\rm rel}'\Theta\!\left(-\bls{n}\cdot\bls{v}_{\rm rel}'\right)\label{particlegasrate},
\end{equation}
where $\bls{n}=d{\bf A}/dA$ denotes the normal vector of the surface element and $n_g$ the gas density. The Heaviside function $\Theta(\cdot)$ ensures that only incident gas particles contribute. Integrating the rate \eqref{particlegasrate} over the surface $\partial V$ of the particle gives the total rate of scattering events as a function of ${\bf p'}$, ${\bf P'}$, and ${\bf J}'$.

The effect of a single collision on the joint state of particle and gas can be expressed by the conditional probability density
\begin{eqnarray} \label{momtrans1}
Q(\bls{P},\bls{J}&&,\bls{p}|\bls{P}',\bls{J}',\bls{p}')=\delta(\bls{P}-\bls{P}'+\bls{p}-\bls{p}')\notag\\
&&\times\delta[\bls{J}-\bls{J}'+\bls{r}\times(\bls{p}-\bls{p}')]\, u\left(\bls{p}\left|\bls{p}',\bls{P}',\bls{J}'\right.\right),
\end{eqnarray}
describing the probability of obtaining the momenta $\bls{p}$, $\bls{P}$ and $\bls{J}$ if they were $\bls{p}'$, $\bls{P}'$ and $\bls{J}'$ prior to the collision. Here we dropped the dependence on ${\bf r}$, ${\bf n}$, and $\Omega$ for brevity. 
The delta functions in Eq.~\eqref{momtrans1} take into account the conservation of total linear and angular momentum, and the function $u\left(\bls{p}\left|\bls{p}',\bls{P}',\bls{J}'\right.\right)$ describes the probability that the gas atom leaves the particle with momentum ${\bf p}$. 

It follows from the principle of relativity that the latter can only depend on the relative momenta before and after the collision, i.e.
\begin{equation} \label{momtrans2}
u\left(\bls{p}\left|\bls{p}',\bls{P}',\bls{J}'\right.\right)=g\left(\bls{q}|\bls{q}'\right),
\end{equation}
where $\bls{q}=\bls{p}\!-\!m\bls{P}'/M\!+\!m\bls{r}\!\times\! \rI^{-1}(\Omega)\bls{J}'$ and $\bls{q}'=\bls{p}'\!-\!m\bls{P}'/M\!+\!m\bls{r}\!\times\! \rI^{-1}(\Omega)\bls{J}'$. 

The transfer function $g({\bf q}|{\bf q}')$ thus  contains all the details of the scattering process. It provides the distribution of outgoing atom momenta ${\bf q}$ given an incident momentum ${\bf q}'$, in the frame of reference where the position of impact ${\bf r}$ is at rest immediately before the collision. 

\subsection{Collision integral}

We are now in the position to express the collision integral in terms of the rate \eqref{particlegasrate} and the change in linear and angular momentum \eqref{momtrans1}. Exploiting that the particle is uncorrelated with the state of the impinging atom one can write the particle state after the infinitesimal time $dt$ as
\begin{align}
f_{t+dt}(\bls{P},\bls{J})=&\int d^3\!p\,d^3\!p'd^3\!P'd^3\!J' f_t(\bls{P}',\bls{J}')\mu(\bls{p}')\, \nonumber \\
&\times w_{dt}(\bls{P}\!,\bls{J},\bls{p}|\bls{P}'\!\!,\bls{J}',\bls{p}'),\label{collintftplusdt}
\end{align}
where we introduced the conditional probability density $w_{dt}$ for having $\bls{p}$, $\bls{P}$ and $\bls{J}$ after the time $dt$ with initial momenta $\bls{p}'$, $\bls{P}'$ and $\bls{J}'$.
We omitted the dependence on ${\bf R}$ and $\Omega$ for brevity. 

The probability for the number of independent collisions in an impact region $dA$ is given by the Poisson distribution with mean $d\Gamma dt$. Assuming that at most  a single collision occurs within $dt$ and
taking into account that every surface element is hit independently one obtains
\begin{align}
&w_{dt}(\bls{P},\bls{J},\bls{p}|\bls{P}',\bls{J}',\bls{p}')=\!\int_{\partial V}\!d\Gamma dt Q(\bls{P},\bls{J},\bls{p}|\bls{P}',\bls{J}',\bls{p}') \nonumber \\
& + \left[1-\int_{\partial V}d\Gamma dt\right]\delta( {\bf p}- {\bf p}')\delta(\bls{P}\!-\!\bls{P}')\delta(\bls{J}\!-\!\bls{J}').\label{collintprobdtlang}
\end{align}
The first term accounts for a single collision in $dt$, while the second describes that no collision occurs.

Inserting \eqref{collintprobdtlang} into \eqref{collintftplusdt}, exploiting the normalization of $\mu$ and $Q$ and drawing the limit $dt \to 0$ yields the rotranslational  collision integral
\begin{align}
\partial_t^{\rm coll}f_t(\bls{P},\bls{J})=&\int d^3\!P'd^3\!J' \left[K(\bls{P},\bls{J}|\bls{P}',\bls{J}')f_t(\bls{P}',\bls{J}')\right.\nonumber \\
&-\left.K(\bls{P}',\bls{J}'|\bls{P},\bls{J})f_t(\bls{P},\bls{J})\right]\,,\label{collintlinboltzeq}
\end{align}
involving the rate densities 
$K(\bls{P},\bls{J}|\bls{P}',\bls{J}') =  \int d^3\!p\, d^3\!p' \int_{\partial V}d\Gamma \mu(\bls{p}') Q(\bls{P},\bls{J},\bls{p}|\bls{P}',\bls{J}',\bls{p}').$
The latter are determined by the transfer function $g({\bf q}|{\bf q}')$ and the momentum distribution  $\mu({\bf p})$ of the gas,
\begin{align} \label{collintprobcurrent}
K(\bls{P},\bls{J}|\bls{P}',\bls{J}') = & \int d^3\!q\, d^3\!q' \int_{\partial V}d\Gamma\, \delta(\bls{P}-\bls{P}'+\bls{q}-\bls{q}') \nonumber \\
&\times\delta[\bls{J}-\bls{J}'+\bls{r}\times(\bls{q}-\bls{q}')]\, g({\bf q}|{\bf q}') \nonumber\\
&\times\mu\Big[\bls{q}'+\frac{m}{M}{\bf P}'-m {\bf r}\times \rI^{-1}(\Omega){\bf J}'\Big]
\end{align}
with
\begin{equation}
d\Gamma=-\frac{n_{\rm g}}{m}d{\bf A}\cdot\bls{q}'\Theta\!\left(-\bls{n}\cdot\bls{q}'\right)\,.
\end{equation}
The first term of the collision integral (\ref{collintlinboltzeq}) describes how the nanoparticle state $f_t$ changes due to the collision-induced probability flow from $({\bf P}', {\bf J}')$ to $(\bls{P},\bls{J})$, while the reverse process is given by the second term.

\subsection{Gas-induced force and torque}

In absence of an external potential, the influence of the gas molecules on the mean momentum $\braket{\bls{P}}$ and mean angular momentum $\braket{\bls{J}}$ follows from \eqref{eq:BE} as
\begin{subequations} \label{fnt}
\begin{equation}
\frac{d}{dt}\braket{\bls{P}}=\int d^3\!Rd^3\Omega d^3\!Pd^3\!J\, \bls{P}\,\partial_t^{\rm coll} f_t(\bls{P},\bls{J}),
\end{equation}
and
\begin{equation}
\frac{d}{dt}\braket{\bls{J}}=\int d^3\!Rd^3\Omega d^3\!Pd^3\!J\, \bls{J}\,\partial_t^{\rm coll} f_t(\bls{P},\bls{J}).
\end{equation}
\end{subequations}

Inserting Eqs.~\eqref{collintlinboltzeq} and \eqref{collintprobcurrent} and carrying out the integration yields the gas-induced,  nonconservative force and torque,
\begin{subequations}\label{fnt2}
\begin{align} \label{momflow}
 \frac{d}{dt}\braket{\bls{P}}= & \left \langle \int d^3q d^3q' \int_{\partial V} d \Gamma\, \left ( {\bf q}' - {\bf q} \right ) g({\bf q}| {\bf q}') \right. \nonumber \\
  & \left. \times \mu \left [ q' + \frac{m}{M} {\bf P} - m {\bf r} \times \rI^{-1}(\Omega){\bf J} \right ] \vphantom{\int}\right \rangle ,
\end{align}
and
\begin{align} \label{angmomflow}
 \frac{d}{dt}\braket{\bls{J}}= & \left \langle \int d^3q d^3q' \int_{\partial V} d \Gamma\, {\bf r} \times \left ( {\bf q}' - {\bf q} \right )  g({\bf q}| {\bf q}')  \right. \nonumber \\
  & \left. \times \mu \left [ q' + \frac{m}{M} {\bf P} - m {\bf r} \times \rI^{-1}(\Omega){\bf J} \right ] \vphantom{\int}\right \rangle .
\end{align}
\end{subequations}
Equation~\eqref{momflow} describes how the balance between the gas momentum flow into and out of the particle surface results in a net force. In a similar fashion, the angular momentum flow through the surface induces the nonconservative torque \eqref{angmomflow}.

The formulas \eqref{momflow} and \eqref{angmomflow} will be used in Sec.~\ref{sec:snd} to calculate the gas-induced force and torque for the special cases of specular and diffuse reflection. Before that, we derive the Fokker-Planck equation by approximating the Boltzmann equation \eqref{collintlinboltzeq} for small particle velocities and momentum kicks.

\section{Rotranslational Fokker-Planck equation} \label{sec:fpe}

In order to derive a Fokker-Planck equation from the rotranslational collision integral \eqref{collintlinboltzeq} we exploit that the particle is much heavier than a gas atom so that we can expand the delta functions in  Eq.~\eqref{collintprobcurrent} for small momentum transfers $\Delta {\bf p}={\bf q}-{\bf q}'$. In doing so we assume that the linear and the angular momentum of the particle is much greater than those transferred in a single collision.

The expansion for small $\Delta {\bf p}$ is carried out conveniently by using that
\begin{subequations}
\begin{align}
 \delta({\bf P} - {\bf P}' + \Delta {\bf p}) = & \left [ 1 + \Delta {\bf p} \cdot \nabla_{\bf P} + \frac{1}{2} \left (\Delta {\bf p} \cdot \nabla_{\bf P} \right )^2 \right ] \nonumber \\ 
 & \times \delta ( {\bf P} - {\bf P}' ),
\end{align}
and
\begin{align}
 \delta({\bf J} - {\bf J}' + {\bf r} \times \Delta {\bf p}) = & \Big [ 1 + {\bf r} \times \Delta {\bf p} \cdot \nabla_{\bf J}   \\
  & + \frac{1}{2} \left ({\bf r} \times \Delta {\bf p} \cdot \nabla_{\bf J} \right )^2 \Big ] \delta({\bf J} - {\bf J}'), \nonumber
\end{align}
\end{subequations}
where the differential operators act on the delta distributions. Using these relations in \eqref{collintprobcurrent} yields, after integrating the delta distributions,
\begin{align}
&\partial_t^{\rm coll}f_t(\bls{P},\bls{J})=\int d^3\!q\, d^3\!q' \int_{\partial V}d\Gamma\,g(\bls{q}|\bls{q}')\nonumber \\
&\times \Big[\Delta\bls{p}\cdot\nabla_{\bls{P}}+\bls{r}\times\Delta\bls{p}\cdot\nabla_{\bls{J}}  + \frac{1}{2} \Big (\Delta\bls{p}\cdot\nabla_{\bls{P}}+\bls{r}\!\times\!\Delta\bls{p}\cdot\nabla_{\bls{J}}\Big)^2 \Big ] \nonumber \\
&\times\mu\!\left[\bls{q}'\!+\!\frac{m}{M}\bls{P}\!-\!m\bls{r}\!\times\!\rI^{-1}(\Omega)\bls{J}\right]f_t(\bls{P},\bls{J}).\label{derivationtwointsdiffapprox}
\end{align}
Here, the derivative operators act on the product of $\mu$ and $f_t$.

The gas distribution function $\mu$ in Eq.~\eqref{derivationtwointsdiffapprox} is shifted by the velocity of the surface element  
multiplied by the atomic mass $m$. Typically, this momentum is much smaller than the width $\sigma_\mu$ of $\mu$ and one can thus expand
\begin{align}
\mu\Big( \bls{q}' + \frac{m}{M}& \bls{P}\!-\!m\bls{r}\!\times\!\rI^{-1}(\Omega)\bls{J}\Big) \approx \mu({\bf q}') \nonumber \\
&  + \left (\frac{m}{M}\bls{P}-m\bls{r}\times\rI^{-1}(\Omega)\bls{J}\right) \cdot \nabla_{\bf q'} \mu({\bf q}'). \label{derivationapproxmu}
\end{align}
Specifically, this approximation is valid if the nanoparticle is close to equilibrium with a thermal gas,
$\sigma_\mu=\sqrt{2mk_{\rm B}T}$,
since $m/M \ll 1$ implies that $\left|m \bls{P}/M-m\bls{r}\times \rI^{-1}(\Omega)\bls{J}\right|/ \sigma_\mu \ll 1$.

Inserting the expansion \eqref{derivationapproxmu} into \eqref{derivationtwointsdiffapprox} and keeping terms up to second order yields the rotranslational Fokker-Planck equation
\begin{widetext}
\begin{equation}
\begin{split}
\partial_t^{\rm coll}f_t=-\left(\begin{array}{c} \!\nabla_{\bls{P}}\! \\ \!\nabla_{\bls{J}}\!\end{array}\right)\!\cdot\!\!\left[{\sf f}(\Omega)-{\sf \Gamma}(\Omega)\left(\begin{array}{c} \!\bls{P}\! \\\! \bls{J}\!\end{array}\right)\right]f_t\!+\!\left(\begin{array}{c} \!\nabla_{\bls{P}}\! \\ \!\nabla_{\bls{J}}\!\end{array}\right)\cdot {\sf D}(\Omega) \left(\begin{array}{c} \!\nabla_{\bls{P}} \!\\\! \nabla_{\bls{J}}\!\end{array}\right)f_t.
\end{split}\label{fokkerplanckgleichung}
\end{equation}
Here we defined the nonconservative force-torque vector
\begin{subequations} \label{derivationpara}
 \begin{equation}
{\sf f}(\Omega) = - \int d^3\!q\, d^3\!q' \int_{\partial V}d\Gamma \mu({\bf q}')g(\bls{q},\bls{q}')\left(\begin{array}{c} \!\bls{q}-\bls{q}'\! \\ \!\bls{r}\times\left(\bls{q}-\bls{q}'\right)\!\end{array}\right)\label{derivationdriftgeneral}, 
\end{equation}
and the rotranslational friction and diffusion tensors
\begin{eqnarray}
\mathsf{\Gamma}(\Omega)&=&m\int d^3\!q\, d^3\!q' \int_{\partial V}d\Gamma g(\bls{q},\bls{q}')\left(\begin{array}{c} \!\bls{q}-\bls{q}'\! \\ \!\bls{r}\times\left(\bls{q}-\bls{q}'\right)\!\end{array}\right)\otimes\left(\begin{array}{c} \!\nabla_{{\bf q}'} \mu({\bf q}')\! \\ \!\bls{r}\times\nabla_{{\bf q}'} \mu({\bf q}')\!\end{array}\right)\mathsf{T}^{-1}(\Omega),\label{derivationgammageneral}\\
\mathsf{D}(\Omega)&=&\frac{1}{2}\int d^3\!q\, d^3\!q' \int_{\partial V}d\Gamma  \mu({\bf q}')g(\bls{q},\bls{q}')\left(\begin{array}{c} \!\bls{q}-\bls{q}'\! \\ \!\bls{r}\times\left(\bls{q}-\bls{q}'\right)\!\end{array}\right)\otimes\left(\begin{array}{c} \!\bls{q}-\bls{q}'\! \\ \!\bls{r}\times\left(\bls{q}-\bls{q}'\right)\!\end{array}\right),\label{derivationdiffusiongeneral}
\end{eqnarray}
\end{subequations}
\end{widetext}
with inertia
\begin{equation}
\mathsf{T}(\Omega)=\left(\begin{array}{rr} M\mathbb{1} & \mathbb{0} \\
     \mathbb{0}\,\,\,\, & \rI(\Omega)\end{array}\right).\label{derivationinertiamatrix}
\end{equation}
We use sans-serif characters to denote  compound vectors and tensors made up of linear and angular momentum components. They allow us to write  Eq.~(\ref{fokkerplanckgleichung}) in compact form.

The Fokker-Planck equation \eqref{fokkerplanckgleichung} describes the coupled rotranslational dynamics of an arbitrarily shaped convex nanoparticle due to a general interaction with the surrounding gas. The gas-induced nonconservative force and torque  \eqref{derivationdriftgeneral} result from a nonvanishing mean momentum flow into or out of the surface of a particle at rest; see Eqs.~\eqref{fnt2}. Moreover, the friction matrix \eqref{derivationgammageneral} is consistent with the expansion of Eqs.~\eqref{fnt2} up to first order in ${\bf P}$ and ${\bf J}$, and thus describes how these momentum flows change due to the particle motion. Finally, the diffusion tensor  Eq.~\eqref{derivationdiffusiongeneral} quantifies the fluctuations of these momentum flows. Note that all three quantities \eqref{derivationpara} depend on the particle orientation $\Omega$, which enters through the scattering rate \eqref{particlegasrate}, the momentum transfer function \eqref{momtrans2}, and the surface integral over $\partial V$.

The Fokker-Planck equation \eqref{fokkerplanckgleichung} reduces to the version derived in Ref.~\cite{galkin2008} for the case of specular and diffuse reflection of thermal gas atoms if the particle is azimuthal and inversion symmetric (point group D$_{\infty{\rm h}}$) and has a constant surface temperature.

\subsection{Friction, diffusion, and equilibration}

In absence of external potentials, the mean change of linear and angular momentum of the particle follow from the Fokker-Planck equation \eqref{fokkerplanckgleichung} as
\begin{equation}
\frac{d}{dt}\left\langle\left(\begin{array}{c} \bls{P} \\\bls{J} \end{array}\right)\right\rangle =\left\langle{\sf f}(\Omega)\right\rangle -\left\langle\mathsf{\Gamma}(\Omega)\left(\begin{array}{c} \bls{P} \\\bls{J} \end{array}\right)\right\rangle .\label{frictiondiffusionerstesmoment}
\end{equation}
Hence, the force and torque described by the vector ${\sf f}(\Omega)$ depends only on the orientation, while those given by the tensor ${\sf \Gamma}(\Omega)$ are linear in the velocities, as characteristic for Stokes friction. Note that ${\sf \Gamma}(\Omega)$  will lead to a gas-induced coupling between the center-of-mass and the rotational motion if it is not block diagonal.

The diffusion tensor  ${\sf D}(\Omega)$ comes into play when considering the expectation value of the kinetic energy,
\begin{align} 
E_{\rm kin}=\frac{1}{2}  \left(\begin{array}{c}\!\! \bls{P}\!\! \\\!\!\bls{J}\!\! \end{array}\right)\cdot\mathsf{T}^{-1}(\Omega)\left(\begin{array}{c} \!\!\bls{P}\!\! \\\!\!\bls{J}\!\! \end{array}\right)\,.
\end{align}
Using  \eqref{fokkerplanckgleichung} one readily finds
\begin{align} 
\frac{d}{dt} E_{\rm kin}=&\left\langle\mathsf{T}^{-1}(\Omega)  {\sf f}(\Omega)\cdot \left(\begin{array}{c}\!\! \bls{P}\!\! \\\!\!\bls{J}\!\! \end{array}\right) \right\rangle 
\nonumber \\
& -\left\langle\left(\begin{array}{c}\!\! \bls{P}\!\! \\\!\!\bls{J}\!\! \end{array}\right)\cdot\mathsf{T}^{-1}(\Omega)  \mathsf{\Gamma}(\Omega)\left(\begin{array}{c} \!\!\bls{P}\!\! \\\!\!\bls{J}\!\! \end{array}\right) \right\rangle 
\nonumber \\
&+\left\langle{\rm Tr}\left[\mathsf{T}^{-1}(\Omega)\mathsf{D}(\Omega)\right]\right\rangle  \,,
\end{align}
where ${\rm Tr}[\cdot]$ denotes the matrix trace. 
The term involving ${\sf \Gamma}(\Omega)$ decreases the energy (or leaves it constant), and thus describes friction if ${\sf \Gamma}(\Omega) {\sf T}(\Omega)$ is positive semidefinite. It will be shown below that this is indeed the case for specular and diffuse reflection of thermally distributed gas atoms. The third term, on the other hand, accounts for the energy increase associated with diffusive motion. It
is non-negative since ${\sf D}(\Omega)$ is positive semidefinite; see Eq.~\eqref{derivationdiffusiongeneral}.

To make explicit that ${\sf D}(\Omega)$  describes diffusion, consider  Eq.~\eqref{fokkerplanckgleichung} for ${\sf f}(\Omega)=0$, ${\sf \Gamma}(\Omega)=0$.  The second moments of momentum then increase linearly with time, as given by the expectation value of ${\sf D}(\Omega)$,  
\begin{align} 
\frac{d}{dt}\left\langle\left(\begin{array}{c}\!\! \bls{P}\!\! \\\!\!\bls{J}\!\! \end{array}\right)\otimes\left(\begin{array}{c} \!\!\bls{P}\!\! \\\!\!\bls{J}\!\! \end{array}\right) \right\rangle  = 2 \langle {\sf D}(\Omega)\rangle  .
\end{align}

Finally, we determine the stationary state $f_{\rm st}$  of the rotranslational Fokker-Planck operator \eqref{fokkerplanckgleichung}. The condition $\partial_t^{\rm coll} f_{\rm st} = 0$ implies that the probability current 
\begin{equation}
 {\sf j} =  \left[{\sf f}(\Omega)-{\sf \Gamma}(\Omega)\left(\begin{array}{c} \!\bls{P}\! \\\! \bls{J}\!\end{array}\right)\right]f_{\rm st} - {\sf D}(\Omega) \left(\begin{array}{c} \!\nabla_{\bls{P}} \!\\\! \nabla_{\bls{J}}\!\end{array}\right)f_{\rm st},
\label{wkstrom}
\end{equation}
must be the curl of a vector field or vanish everywhere \cite{gardiner1985}. It can be demonstrated \cite{gardiner1985,risken1996} that a solution with ${\sf j}=0$ exists only if  ${\sf D}^{-1}(\Omega) {\sf \Gamma}(\Omega)$ is symmetric, yielding
\begin{align} \label{statsol}
 f_{\rm st} \propto & \exp \left \{ -\frac{1}{2} \left [ \mathsf{\Gamma}^{-1} {\sf f}-\left(\begin{array}{c} \bls{P} \\\bls{J} \end{array}\right) \right] \cdot {\sf D}^{-1} {\sf \Gamma}  \left [\mathsf{\Gamma}^{-1} {\sf f}-\left(\begin{array}{c} \bls{P} \\\bls{J} \end{array}\right) \right] \right \},
\end{align}
where we dropped the orientation dependence for brevity. This solution is then unique and every initial state approaches it asymptotically \cite{gardiner1985}. The dependence of $f_{\rm st}$ on ${\bf R}$ and $\Omega$ then  follows from $\partial_t^{\rm cons}f_{\rm st} = 0$, as discussed in Sect.~\ref{sec:snd} and  App.~\ref{app:therm} for specular and diffuse reflection.

\subsection{Langevin equations}

The dynamics described by the Fokker-Planck equation can be equivalently represented by a stochastic Langevin equation \cite{gardiner1985}. Including the conservative force and torque ${\sf f}_{\rm c}({\bf R},\Omega)$, the  Langevin equation for linear and angular momentum reads
\begin{align} \label{langevin}
 \left(\begin{array}{c} d\bls{P} \\ d \bls{J} \end{array}\right) = & \left [{\sf f}_{\rm c}({\bf R},\Omega) + {\sf f}(\Omega) - \mathsf{\Gamma}(\Omega)\left(\begin{array}{c} \bls{P} \\\bls{J} \end{array}\right)  \right ] dt \nonumber \\
  & + \sqrt{2 {\sf D}(\Omega)} d{\sf W}_t,
\end{align}
where $d{\sf W}_t$ is a vector of independent Wiener increments. Since this equation depends on the position and orientation of the nanoparticle, it has to be supplemented by the kinematic equations for position and rotation matrix,
\begin{equation} \label{kin}
 d {\bf R} = \frac{ {\bf P}}{M} dt \quad {\rm and} \quad d \rR(\Omega) = \rI^{-1}(\Omega) {\bf J} \times \rR(\Omega) dt.
\end{equation}
Numerically, it is often much more convenient to solve the coupled stochastic differential equations \eqref{langevin} and \eqref{kin} instead of a partial differential equation for the probability density. They will be used in Sec.~\ref{sec:snd} to illustrate the phase space dynamics of the linear rigid rotor.

\section{Specular and Diffuse Reflection} \label{sec:snd}

In this section we will specify the friction and diffusion tensors for the special cases of specular and diffuse reflection. The latter are phenomenological descriptions of surface scattering, frequently employed to avoid dealing with the complexity of atomistically exact interactions \cite{cercignani1988}.

In the simple case of specular reflection the atoms are elastically reflected from the particle surface, just as light from a perfect mirror. Such a description can 
even be used  to approximately describe atom-molecule scattering in so-called rigid shell models \cite{z1,z2,z3,z4}.
The model of diffuse reflection, on the other hand, accounts for the fact that atoms get adsorbed and reemitted by the surface. Their final momentum is then determined by the surface temperature $T_{\rm s}$. Introducing the accommodation coefficient $\alpha_{\rm c} \in [0,1]$ allows one to continuously interpolate between these two scenarios, with $\alpha_{\rm c} = 0$ referring to specular reflection. For both reflection types the momentum transfer \eqref{momtrans2} can be given explicitly.

\begin{figure*}[]
\centering
\includegraphics[width = 0.99\textwidth]{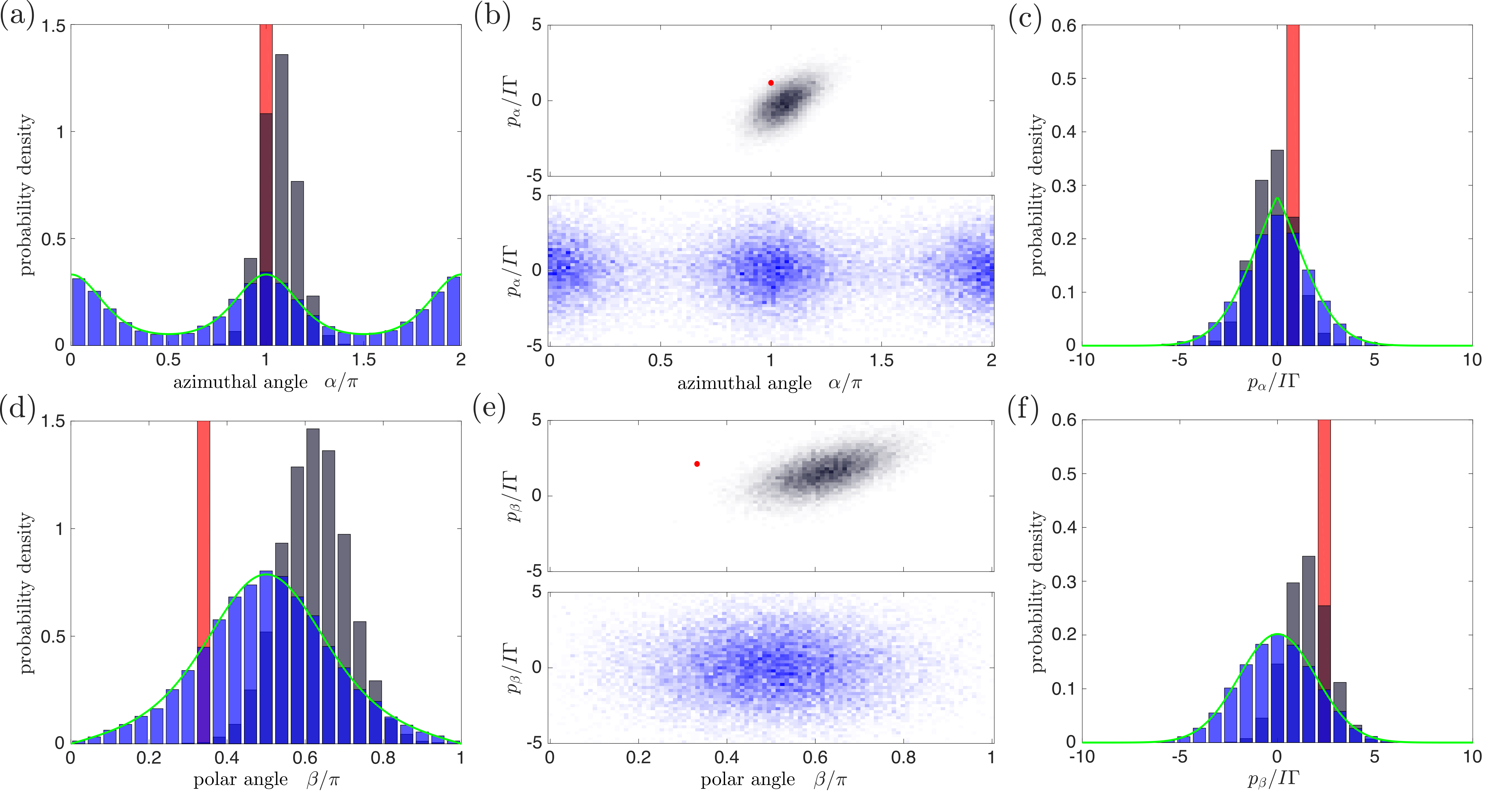}
	\caption{(Color online) Thermalization dynamics of the linear rigid rotor. The panels (a)-(f) show different marginals of the four-dimensional phase space distribution function at three different instances of time, in red, gray, and blue. The initial state, indicated in red, is given by $(\alpha,\beta,p_\alpha,p_\beta)=(\pi,\pi/3,I{\it\Gamma},2 I{\it\Gamma})$, while the time-evolved state at $t = 0.5/{\it \Gamma}$ and $t = 13/{\it \Gamma}$ is shaded in gray and blue, respectively. Panels (a), (c), (d), and (f) show histograms for the individual phase space coordinates, while (b) and (e) are density plots for the reduced azimuthal and polar degrees of freedom.
	Note that the final distribution is well approximated by the thermal state (green lines).	The plots are obtained by calculating several thousand trajectories from the stochastic differential equations \eqref{linlangevin} for $I {\it \Gamma}^2 / k_{\rm B} T = 0.26$ and $V_0 / k_{\rm B} T = 2.42$.  }\label{picsimphasenraum}
\end{figure*}

\subsection{Transfer function}

\subsubsection{Specular Reflection } 
The momentum ${\bf p}$ of a gas atom elastically reflected from the surface element $dA$ at separation ${\bf r}$ from the center of mass of a moving and rotating nonspherical nanoparticle is determined by the conservation of energy, momentum, and angular momentum. Given the initial momenta ${\bf p}'$, ${\bf P}'$, and ${\bf J}'$ and using that the momentum is conserved in the surface plane yields 
\begin{equation}
{\bf p} ={\bf p}'-2\widetilde{m}(\Omega) \bls{n}\left[\frac{{\bf p}'}{m} - \frac{{\bf P}'}{M} + {\bf r} \times \rI^{-1}(\Omega) {\bf J}' \right] \cdot \bls{n}.
\end{equation}
Here, ${\bf n}$ is the surface normal vector and we defined the orientation-dependent effective mass for each surface element
\begin{equation}
\widetilde m(\Omega) = \frac{m M}{m + M +m M (\bls{r}\times\bls{n})\cdot\rI^{-1}(\Omega)\left(\bls{r}\times\bls{n}\right)}.
\end{equation}

For small mass ratios, $m/M \ll 1$, we can approximate $\widetilde{m} \approx m$ and obtain the momentum transfer function of specular reflection as
\begin{equation} \label{momtransspec}
g_{\rm sp}(\bls{q}|\bls{q}')=\delta \left[\bls{q}-\bls{q}'+2\bls{n}\left(\bls{n}\cdot\bls{q}'\right)\right].
\end{equation}
The transferred momentum always points into the direction of the surface normal vector. This implies that the angular momentum transfer in certain directions vanishes for special shapes, such as along the symmetry axis of a cylinder or in all directions for a sphere. This is discussed in App.~\ref{app:therm} and reflected in Tab.~\ref{tab:fric} and \ref{tab:diff}.

\subsubsection{ Diffuse reflection } 
The model of diffuse reflection assumes that the incident gas atom thermalizes with the nanoparticle surface at temperature $T_{\rm s}$ before being thermally reemitted \cite{cercignani1988}; see Fig.~\ref{hauptbild}. The momentum distribution of the ejected gas atoms therefore reads as
\begin{equation}
g_{\rm di}(\bls{q}|\bls{q}')=\frac{\bls{n}\cdot\bls{q}\,}{2\pi \left(mk_{\rm B}T_{\rm s}\right)^2}{\rm exp}\left(-\frac{\bls{q}^2}{2mk_{\rm B}T_{\rm s}}\right)\Theta(\bls{n}\cdot\bls{q}),
\end{equation}
independent of the incident momentum $\bls{q}'$. We will see that for constant $T_{\rm s}$ this transfer function leads to thermalization independent of the nanoparticle shape. If the surface temperature is a function of the point of impact ${\bf r}$ diffuse reflection describes photophoresis \cite{rohatschek1995,zulehner1995}.

To model experimental data, a combination of specular and diffuse reflection is often specified via the fraction of diffusely reflected molecules $\ac$, referred to as \emph{accommodation coefficient},
\begin{equation}
g(\bls{q},\bls{q}')=(1-\ac)g_{\rm sp}(\bls{q},\bls{q}')+\ac g_{\rm di}(\bls{q},\bls{q}').\label{diffreflectiontransfboth}
\end{equation}

\subsection{Gas-induced forces and torques}

The nonconservative force and torque \eqref{fnt2} predicted by the rotranslational Boltzmann equation  are of the form
\begin{subequations}
\begin{eqnarray}
 \frac{d}{dt}\braket{\bls{P}} &=& \left\langle\int_{\partial V}dA\frac{d\bls{F}}{dA}\right\rangle  ,
 \\
 \frac{d}{dt}\braket{\bls{J}} &=& \left\langle\int_{\partial V}dA\,{\bf r} \times \frac{d\bls{F}}{dA}\right\rangle ,
\end{eqnarray} 
\end{subequations}
where the force increment $d {\bf F}$ follows from \eqref{diffreflectiontransfboth} as
\begin{align}
\frac{d {\bf F}}{dA} = 2(1-\ac)\bls{n}\left(\bls{n}\cdot\bls{j}_{\rm m}\right) + \ac\left(\bls{j}_{\rm m}-\frac{j_{\rm a}}{2}\sqrt{2\pi m k_{\rm B}T_{\rm s}}\bls{n}\right).
\end{align}
Here we introduced the incident atom flux
\begin{equation} \label{atflux}
j_{\rm a}=\frac{n_{\rm g}}{m}\int \limits_{\bls{n}\cdot \bls{p}\le 0}d^3p\, |\bls{n}\cdot \bls{p}|\,\mu\!\left(\bls{p}+\frac{m}{M}\bls{P}-m\bls{r}\times\rI^{-1}(\Omega)\bls{J}\right),
\end{equation}
and the corresponding momentum flux
\begin{equation} \label{momflux}
\bls{j}_{\rm m}=\frac{n_{\rm g}}{m}\int \limits_{\bls{n}\cdot \bls{p}\le 0}d^3p\, \bls{p}|\bls{n}\cdot \bls{p}|\,\mu\!\left(\bls{p}+\frac{m}{M}\bls{P}-m \bls{r}\times\rI^{-1}(\Omega)\bls{J}\right).
\end{equation}
For the remainder of this section we take the gas to be thermally distributed,
\begin{equation}\label{eq:Boltzmann}
\mu(\bls{p})=\frac{1}{(\sqrt{2\pi m k_{\rm B}T})^{3}}{\rm exp}\left(-\frac{\bls{p}^2}{2mk_{\rm B}T}\right)\,.
\end{equation}
As in Sec.~\ref{sec:fpe} we assume the velocity of the surface element  $\bls{v}_{dA}=\bls{P}/M-\bls{r}\times\rI^{-1}(\Omega)\bls{J}$ to be much smaller than the most probable gas velocity $\sqrt{2k_{\rm B}T/m}$. We can thus expand the Boltzmann distribution
in \eqref{atflux} and \eqref{momflux} up to second order, yielding
\begin{align}\label{collforceapproxforce}
\frac{d\bls{F}}{dA} = & -\frac{1}{2} n_{\rm g}\ac k_{\rm B}T \gamma_{\rm s} {\bf n}  - n_{\rm g} \sqrt{\frac{m k_{\rm B} T}{2 \pi}}  \vphantom{\frac{1}{2}}\ac\bls{v}_{dA} \nonumber  \\
& - n_{\rm g} \sqrt{\frac{m k_{\rm B} T}{2 \pi}} \left(4-3\ac+\frac{\pi\ac\gamma_{\rm s}}{2}\right)\left(\bls{v}_{dA}\cdot\bls{n}\right)\bls{n}\vphantom{\frac{1}{2}} \nonumber \\
& - \frac{1}{2}n_g \ac m \left(\bls{v}_{dA}\cdot\bls{n}\right)\bls{v}_{dA} \nonumber
\\
& - \frac{1}{4}n_{\rm g} m\left[4(1-\ac)+\ac\gamma_{\rm s}\right](\bls{v}_{dA}\cdot\bls{n})^2\bls{n}.
\end{align}
where $\gamma_{\rm s}=\sqrt{T_{\rm s}/T}$ can be a function of ${\bf r}$.

The first term on the right hand side is a velocity-independent nonconservative force and torque. It can also be obtained from the Fokker-Planck equation by evaluating \eqref{derivationdriftgeneral} for the momentum transfer function \eqref{diffreflectiontransfboth}. Note that it vanishes whenever the surface temperature $T_{\rm s}$ is uniform, while an inhomogeneous surface temperature, $\gamma_{\rm s} \equiv \gamma_{\rm s}({\bf r})$, results in \emph{photophoresis} \cite{rohatschek1995,zulehner1995} with the force and torque 
\begin{equation}
 {\sf f}(\Omega) = -\frac{1}{2}\ac n_{\rm g}k_{\rm B}T\int_{\partial V} dA\,\gamma_{\rm s}({\bf r})\left(\begin{array}{c} \!\!\boldsymbol{n}\!\! \\ \!\!\boldsymbol{r}\times\boldsymbol{n}\!\!\! \end{array}\right).
\end{equation}

The fourth term on the right hand side of Eq.~\eqref{collforceapproxforce} describes the \emph{inverse Magnus effect} \cite{borg2003}. The resulting force and torque can be calculated for particles of homogeneous mass density $\varrho$ and arbitrary shape as
\begin{equation}
  \left ( \begin{array}{c}
 {\bf F}_{\rm iM}\\
 {\bf N}_{\rm iM}
  \end{array} \right )
  = \frac{1}{2} \frac{\ac n_{\rm g} m}{\varrho} \left ( \begin{array}{c}
          \bls{P}\times\rI^{-1}(\Omega)\bls{J} \\
   \bls{J}\times\rI^{-1}(\Omega)\bls{J}
     \end{array} \right ).
\end{equation}
Compared to the conventional Magnus effect, the force on the spinning particle points into the opposite direction. For symmetric tops, the corresponding torque leads to a precession of the angular momentum vector with a constant frequency. The inverse Magnus effect vanishes for pure specular reflection ($\alpha_{\rm c} = 0$).

The second and third term in Eq.~\eqref{collforceapproxforce}, which are linear in ${\bf v}_{ dA}$, yield a rotranslational friction force and torque. They are the same as those implied by the Fokker-Planck equation \eqref{fokkerplanckgleichung}, as obtained by evaluating the friction tensor \eqref{derivationgammageneral} with  \eqref{diffreflectiontransfboth},
\begin{widetext}
\begin{eqnarray}\label{derivationgammaspecular}
\mathsf{\Gamma}(\Omega)&=&n_{\rm g}\sqrt{\frac{mk_{\rm B}T}{2 \pi}}\int_{\partial V} dA \left\{\left[4-3\ac+\frac{\pi \ac\gamma_{\rm s}}{2}\right]\left(\begin{array}{c} \!\!\boldsymbol{n}\!\! \\ \!\!\boldsymbol{r}\times\boldsymbol{n}\!\!\! \end{array}\right)\!\otimes\!\left(\begin{array}{c} \!\!\boldsymbol{n}\!\! \\ \!\!\boldsymbol{r}\times\boldsymbol{n}\!\!\! \end{array}\right)\!+\!\ac\left(\begin{array}{rr} \mathbb{1}\,\, & -\mathsf{r} \\
     \mathsf{r}\,\, & -\mathsf{r}^2\end{array}\right)\right\}\mathsf{T}^{-1}(\Omega).
\end{eqnarray}
Here, the matrix $\mathsf{r}$ is defined such that $\mathsf{r}\bls{a}=\bls{r}\times\bls{a}$ for any vector $\bls{a}$. The associated diffusion tensor follows from \eqref{derivationdiffusiongeneral} as
\begin{eqnarray}\label{derivationdiffusionspecular}
 \mathsf{D}(\Omega)&=&k_{\rm B}Tn_{\rm g}\sqrt{\frac{mk_{\rm B}T}{2 \pi}}\int_{\partial V} dA \left\{\left[4-\frac{7-\gamma_{\rm s}^2}{2}\ac+\frac{\pi\ac\gamma_{\rm s}}{2}\right]\left(\begin{array}{c} \!\!\boldsymbol{n}\!\! \\ \!\!\boldsymbol{r}\times\boldsymbol{n}\!\!\! \end{array}\right)\!\otimes\!\left(\begin{array}{c} \!\!\boldsymbol{n}\!\! \\ \!\!\boldsymbol{r}\times\boldsymbol{n}\!\!\! \end{array}\right)\!+\!\frac{1+\gamma_{\rm s}^2}{2}\ac\left(\begin{array}{rr} \mathbb{1}\,\, & -\mathsf{r} \\
     \mathsf{r}\,\, & -\mathsf{r}^2\end{array}\right)\right\}.
\end{eqnarray}
\end{widetext}

For $\gamma_{\rm s} = 1$, the friction tensor \eqref{derivationgammaspecular} and the diffusion tensor \eqref{derivationdiffusionspecular} obey the relation
\begin{equation}
\mathsf{D}(\Omega)=k_{\rm B}T\mathsf{\Gamma}(\Omega)\mathsf{T}(\Omega), \label{dissipationfluctrelation}
\end{equation}
which is reminiscent of the fluctuation-dissipation relation. It implies that the Fokker-Planck operator \eqref{fokkerplanckgleichung} admits the stationary solution \eqref{statsol} with ${\sf f}(\Omega) = 0$. The additional requirement that the stationary solution is also invariant under the conservative time evolution yields the unique thermal equilibrium state
\begin{align} \label{feq}
f_{\rm eq} = & \frac{\sqrt{g(\Omega)}}{Z} \exp \left [ - \frac{1}{k_{\rm B} T}\left(\frac{{\bf P}^2}{2 M} + \frac{1}{2} {\bf J} \cdot \rI^{-1}(\Omega) {\bf J} \right ) \right ] \nonumber \\
& \times \exp \left [- \frac{V({\bf R}, \Omega)}{k_{\rm B} T} \right ],
\end{align}
where $Z$ is the partition function. Here, $g(\Omega)$ is the metric determinant of the orientational configuration space, and thus also the squared Jacobian determinant when transforming to the phase space (see App.~\ref{app:tdf}). It is demonstrated in App.~\ref{app:therm} that every initial state converges towards this distribution if $\ac \neq 0$.

\subsection{Thermalization of the linear rigid rotor}\label{sec:trl}

We apply the relations derived above to study the rotational thermalization dynamics of a linear rigid rotor of length $\ell \gg R$ and moment of inertia $I = M \ell^2/12$. The form of its rotational friction and diffusion tensors is given in Tabs.~\ref{tab:fric} and \ref{tab:diff} for arbitrary accommodation coefficients $\alpha_{\rm c}$ and surface temperatures $T_{\rm s}$. Here, we make use of this result for $T_{\rm s} = T$ in order to express the Fokker-Planck equation in terms of the phase space distribution $h_t(\alpha,\beta,p_\alpha,p_\beta)$ of the orientation state (see App.~\ref{app:tdf}),
\begin{align}
\partial_t h_t + \{h_t, H \} = & {\it \Gamma}\left[\partial_{p_\alpha}\left(p_\alpha h_t\right) +  \partial_{p_\beta}\left(p_\beta h_t\right)\right ] \nonumber \\
 & +D \left (\sin^2 \beta \partial^2_{p_\alpha}h_t + \partial^2_{p_\beta}h_t\right)\,.\label{simfokpllinplanrot}
\end{align}
Here we use the Poisson bracket $\{ \cdot , H \}$ with the Hamilton function
\begin{equation} \label{ham}
 H = \frac{1}{2 I} \left ( \frac{p_\alpha^2}{\sin^2 \beta} + p_\beta^2 \right ) - V_0 \cos^2 \alpha \sin^2 \beta.
\end{equation}
Note that \eqref{dissipationfluctrelation} reduces to $D = k_{\rm B} T {\it \Gamma} I$.

The angles $\alpha \in [0,2 \pi]$ and $\beta \in [0,\pi]$ specify the orientation of the symmetry axis ${\bf m}$; see Tab.~\ref{tab:fric}, while $p_\alpha$ and $p_\beta$ are the corresponding conjugate angular momenta. The potential energy in \eqref{ham} describes for instance the  interaction between a polarizable rod-shaped particle and the field of a laser beam polarized in $x$ direction \cite{stapelfeldt2003}.

According to Eq.~\eqref{langevin}, the dynamics described by \eqref{simfokpllinplanrot} can also be expressed in terms of the set of stochastic differential equations,
\begin{subequations} \label{linlangevin}
\begin{align}
d \alpha & = \partial_{p_\alpha}  H dt, \\
d \beta & = \partial_{p_\beta}  H dt, \\
d p_\alpha & = - \partial_\alpha H dt - {\it \Gamma} p_\alpha dt + \sqrt{2 D} \sin \beta \,dW^{(\alpha)}_t, \\
d p_\beta & = - \partial_\beta H dt - {\it \Gamma} p_\beta dt + \sqrt{2 D}\, dW_t^{(\beta)}.
\end{align}
\end{subequations}
In Fig.~\ref{picsimphasenraum} we show the thermalization dynamics of a linear rotor with $I {\it \Gamma}^2 / k_{\rm B} T = 0.26$ and $V_0 / k_{\rm B} T = 2.42$, as inspired from recent experiments with levitated silicon nanorods \cite{kuhn2017a}. Based on several thousand trajectories, time evolved according to \eqref{linlangevin}, the figure illustrates how an initially well localized phase space distribution disperses due to diffusion and relaxes towards the thermal state $h_{\rm eq} = \exp ( - H/k_{\rm B} T)/Z$ on the timescale $1/{\it \Gamma}$.

\begin{table*}[tbp]
\caption{Center-of-mass (cm) and rotational (rot) friction tensors of homogeneous spheres, cylinders, and cuboids for specular and diffuse reflection with accommodation coefficient $\ac$ and surface temperature $T_{\rm s} = \gamma_{\rm s}^2 T$. The gas temperature is denoted by $T$, $n_{\rm g}$ is the gas density, $m$ the mass of a gas atom, and $M$ the total mass of the particle. The radius of the sphere and cylinder is $R$, the cylinder length is $\ell$ with symmetry axis ${\bf m}(\Omega)$, and the edge lengths of the cuboid are $a$, $b$, and $c$ with principal axes ${\bf n}_a(\Omega)$, ${\bf n}_b(\Omega)$, and ${\bf n}_c(\Omega)$; see sketches. All friction tensors satisfy the relations \eqref{flucdiss2} with the corresponding diffusion tensors in Tab.~\ref{tab:diff}.}
\begin{center}
\begin{tabular}{|l|}
\hline
\rowcolor[gray]{.8}\parbox{0.98\textwidth}{\centering \textbf{Sphere}} \\ \hline
\parbox{0.18\textwidth}{
\includegraphics[width = 0.1\textwidth]{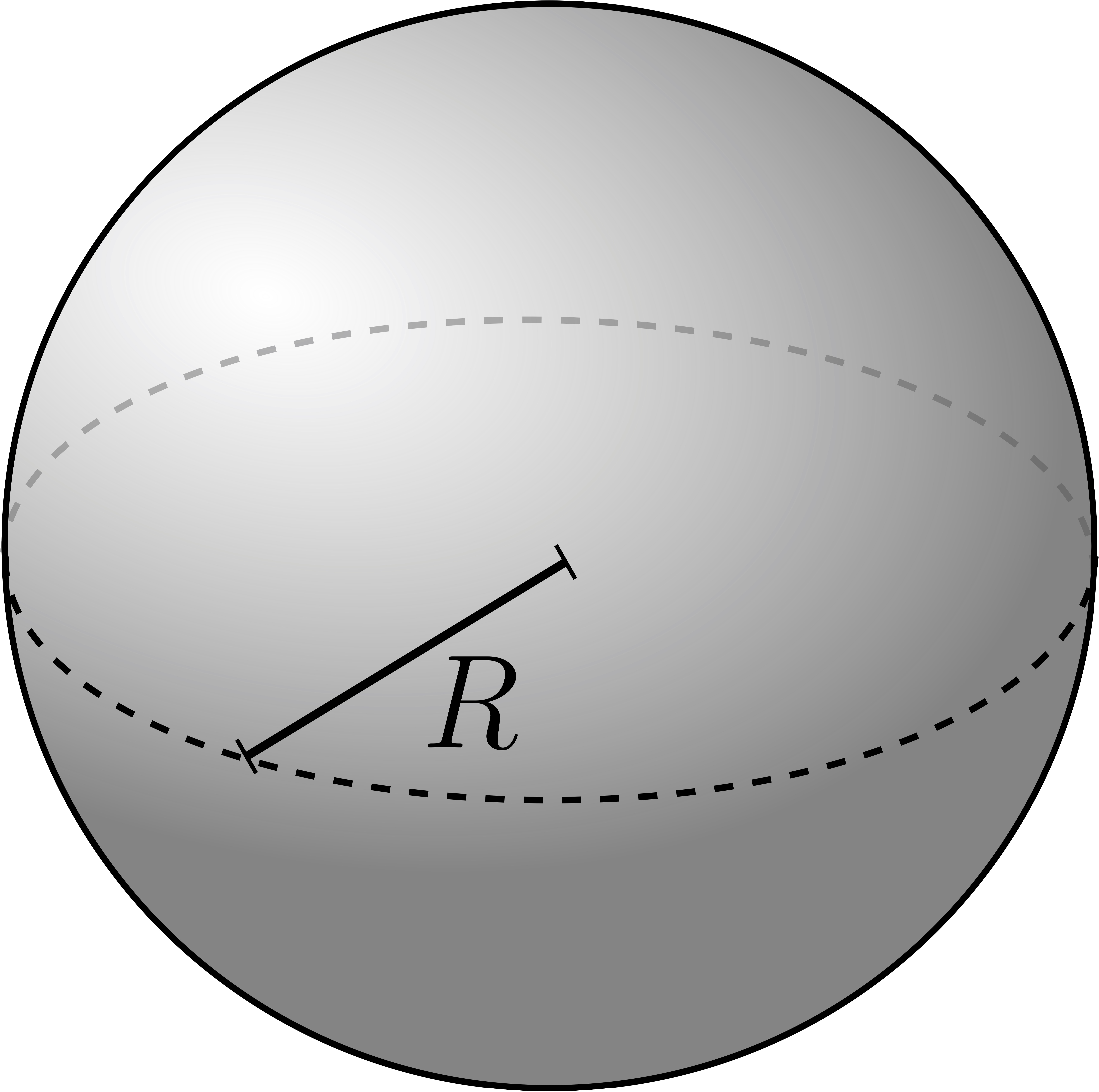}
}
\parbox{0.8\textwidth}{\begin{eqnarray}
\Gamma_{\rm cm}&=&\frac{8n_gR^2\sqrt{2\pi m k_{\rm B} T}}{3M}\left(1+\frac{\pi}{8}\ac\gamma_{\rm s}\right)\mathbb{1}\nonumber\\
\Gamma_{\rm rot}&=&\frac{10\ac n_gR^2\sqrt{2\pi mk_{\rm B}T}}{3M}\mathbb{1}\nonumber
\end{eqnarray}}\\ \hline
\rowcolor[gray]{.8}\parbox{0.98\textwidth}{\centering \textbf{Cylinder}} \\ \hline
\parbox{0.18\textwidth}{
\vspace{0.8mm}
\includegraphics[width = 0.08\textwidth]{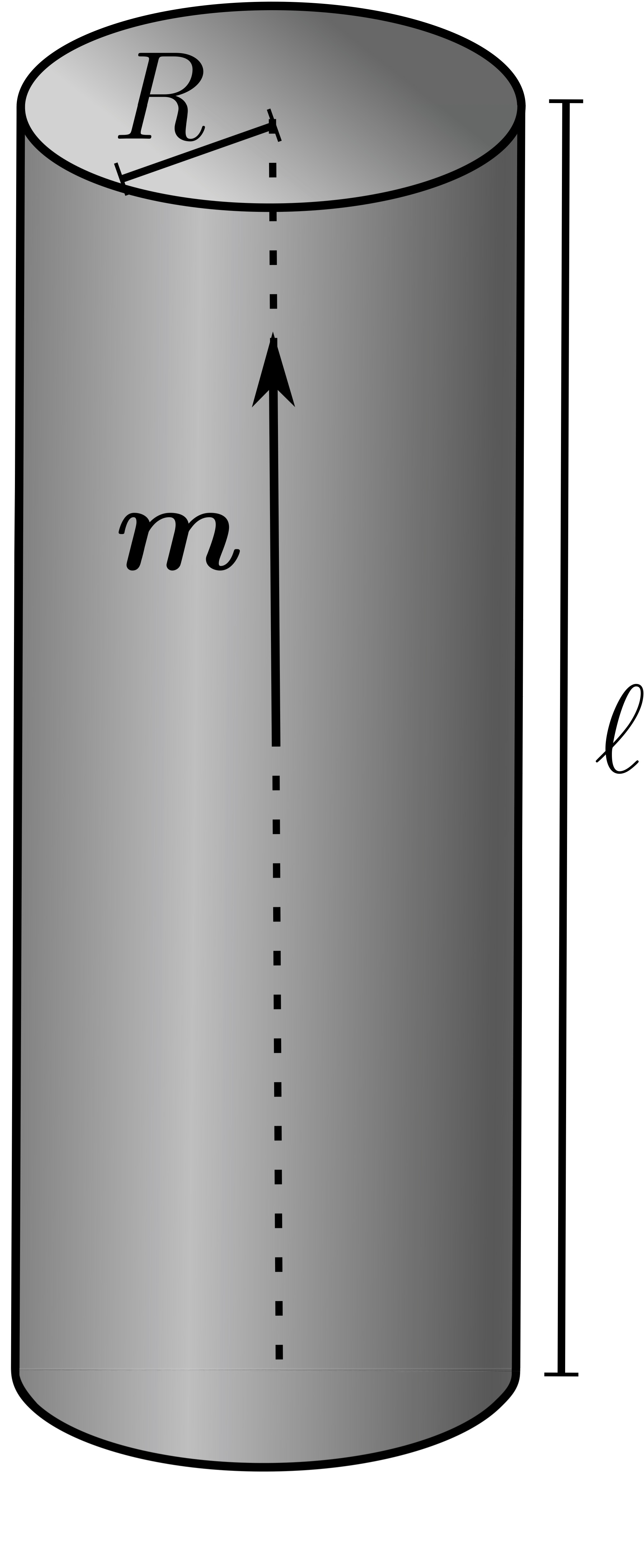}
}
\parbox{0.8\textwidth}{
\begin{eqnarray}
\Gamma_{\rm cm}&=&{\it\Gamma}_{\rm cm}^\perp(\mathbb{1-\bls{m}\otimes\bls{m}})+{\it \Gamma}_{\rm cm}^\parallel\bls{m}\otimes\bls{m}\nonumber\\
\Gamma_{\rm rot}&=&{\it \Gamma}_{\rm rot}^\perp(\mathbb{1-\bls{m}\otimes\bls{m}})+{\it\Gamma}_{\rm rot}^\parallel\bls{m}\otimes\bls{m}\nonumber\\
&&\nonumber\\
{\it\Gamma}_{\rm cm}^\perp&=&\frac{n_gR\ell \sqrt{2\pi m k_{\rm B}T}}{M}\left[2+\ac\left(-\frac{1}{2}+\frac{\pi \gamma_{\rm s}}{4}+\frac{R}{\ell}\right)\right]\nonumber\\
{\it\Gamma}_{\rm cm}^\parallel&=&\frac{n_gR\ell \sqrt{2\pi m k_{\rm B}T}}{M}\left[4\frac{R}{\ell}+\ac\left(1-2\frac{R}{\ell}+\frac{\pi\gamma_{\rm s}}{2}\frac{R}{\ell}\right)\right]\nonumber\\
{\it \Gamma}_{\rm rot}^\perp&=&\frac{n_gR\ell \sqrt{2\pi m k_{\rm B}T}}{M}\frac{\ell^2}{3R^2+\ell^2}\left\{2+12\frac{R^3}{\ell^3}+\ac\left[-\frac{1}{2}+\frac{\pi\gamma_{\rm s}}{4}+3\frac{R}{\ell}+6\frac{R^2}{\ell^2}+\left(\frac{3\pi\gamma_{\rm s}}{2}-6\right)\frac{R^3}{\ell^3}\right]\right\}\nonumber\\
{\it \Gamma}_{\rm rot}^\parallel&=&\frac{n_gR\ell \sqrt{2\pi m k_{\rm B}T}}{M}\ac\left(2+\frac{R}{\ell}\right)\nonumber
\end{eqnarray}
} \\ \hline
\rowcolor[gray]{.8}\parbox{0.98\textwidth}{\centering \textbf{Cuboid}} \\ \hline
\parbox{0.18\textwidth}{
\vspace{0.5mm}
\includegraphics[width = 0.13\textwidth]{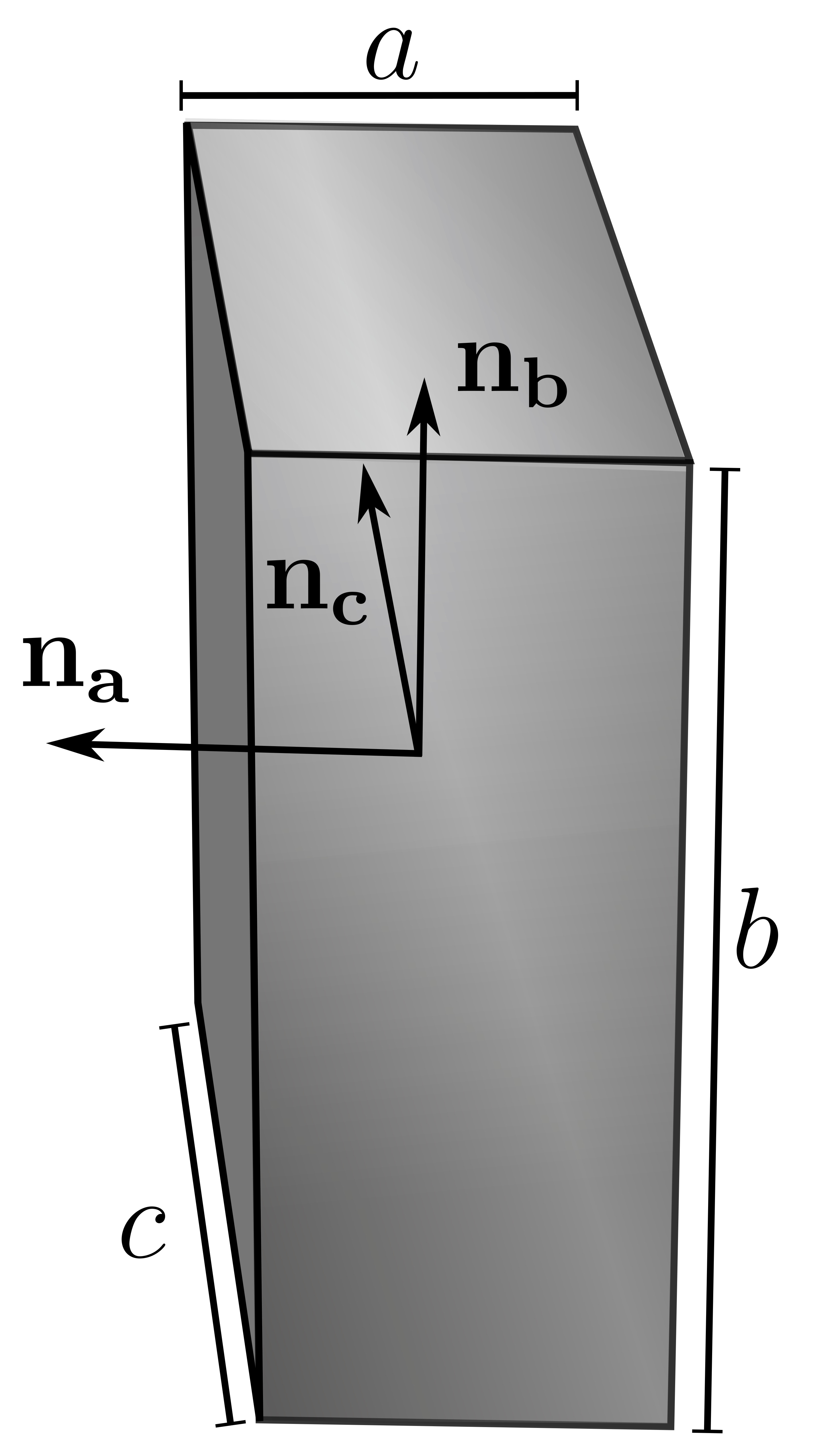}
}
\parbox{0.8\textwidth}{\begin{eqnarray}
\Gamma_{\rm cm}&=&{\it \Gamma}_{\rm cm}^{abc}\bls{n}_a\otimes\bls{n}_a+{\it \Gamma}_{\rm cm}^{bac}\bls{n}_b\otimes\bls{n}_b+{\it \Gamma}_{\rm rot}^{cba}\bls{n}_c\otimes\bls{n}_c\nonumber\\
\Gamma_{\rm rot}&=&{\it \Gamma}_{\rm rot}^{abc}\bls{n}_a\otimes\bls{n}_a+{\it \Gamma}_{\rm rot}^{bac}\bls{n}_b\otimes\bls{n}_b+{\it \Gamma}_{\rm rot}^{cba}\bls{n}_c\otimes\bls{n}_c\nonumber\\
&&\nonumber\\
{\it \Gamma}_{\rm cm}^{abc}&=&\frac{n_gbc \sqrt{2\pi mk_{\rm B}T}}{\pi M}\left[4+\ac\left( -2+\frac{ab+ac}{bc}+\frac{\pi\gamma_{\rm s}}{2}\right)\right]\nonumber\\
{\it \Gamma}_{\rm rot}^{abc}&=&\frac{n_gbc\sqrt{2\pi mk_{\rm B}T}}{\pi M}\!\left\{\frac{4a(c^3+b^3)}{bc\left(b^2+c^2\right)}\!+\ac\!\left[1-\!\left(2-\!\frac{\pi\gamma_{\rm s}}{2}\right)\frac{a(c^3+b^3)}{bc\left(b^2+c^2\right)}+\!\frac{3a(b+c)}{b^2+c^2}\right]\!\right\}\nonumber
\end{eqnarray} 
} \\ \hline
\end{tabular}
\end{center}
\label{tab:fric}
\end{table*}

\begin{table*}[tbp]
\caption{Center-of-mass (cm) and rotational (rot) diffusion tensors of homogeneous spheres, cylinders, and cuboids for specular and diffuse reflection with accommodation coefficient $\ac$ and surface temperature $T_{\rm s} = \gamma_{\rm s}^2 T$. The gas temperature is denoted by $T$, $n_{\rm g}$ is the gas density, $m$ the mass of a gas atom, and $M$ the total mass of the particle. The radius of the sphere and cylinder is $R$, the cylinder length is $\ell$ with symmetry axis ${\bf m}(\Omega)$, and the edge lengths of the cuboid are $a$, $b$, and $c$ with principal axes ${\bf n}_a(\Omega)$, ${\bf n}_b(\Omega)$, and ${\bf n}_c(\Omega)$; see sketches. All diffusion tensors satisfy the relations \eqref{flucdiss2} with the corresponding friction tensors in Tab.~\ref{tab:fric}.}
\begin{center}
\begin{tabular}{|l|}
\hline
\rowcolor[gray]{.8} \parbox{0.98\textwidth}{\centering \textbf{Sphere}}\\ \hline
\parbox{0.18\textwidth}{
\includegraphics[width = 0.1\textwidth]{TableFig1.jpg}
}
\parbox{0.8\textwidth}{\begin{eqnarray}
\mathrm{D}_{\rm cm}&=&\frac{8}{3}n_gR^2\sqrt{2\pi m \left(k_{\rm B} T\right)^3}\left(1+\frac{\pi}{8}\ac\gamma_{\rm s}\right)\mathbb{1}\nonumber\\
\mathrm{D}_{\rm rot}&=&\frac{4}{3}\ac n_gR^4\sqrt{2\pi m\left(k_{\rm B}T\right)^3}\mathbb{1}\nonumber
\end{eqnarray}}\\ \hline
\rowcolor[gray]{.8}\parbox{0.98\textwidth}{\centering \textbf{Cylinder}} \\ \hline
\parbox{0.18\textwidth}{
\vspace{0.8mm}
\includegraphics[width = 0.08\textwidth]{TableFig2.jpg}
}
\parbox{0.8\textwidth}{\begin{eqnarray}
\mathrm{D}_{\rm cm}&=&D_{\rm cm}^\perp(\mathbb{1-\bls{m}\otimes\bls{m}})+D_{\rm cm}^\parallel\bls{m}\otimes\bls{m}\nonumber\\
\mathrm{D}_{\rm rot}&=&D_{\rm rot}^\perp(\mathbb{1-\bls{m}\otimes\bls{m}})+D_{\rm rot}^\parallel\bls{m}\otimes\bls{m}\nonumber\\
&&\nonumber\\
D_{\rm cm}^\perp&=&n_gR\ell \sqrt{2\pi m \left(k_{\rm B}T\right)^3}\left[2+\ac\left(-\frac{1}{2}+\frac{\pi \gamma_{\rm s}}{4}+\frac{R}{\ell}\right)\right]\nonumber\\
D_{\rm cm}^\parallel&=&n_gR\ell \sqrt{2\pi m \left(k_{\rm B}T\right)^3}\left[4\frac{R}{\ell}+\ac\left(1-2\frac{R}{\ell}+\frac{\pi\gamma_{\rm s}}{2}\frac{R}{\ell}\right)\right]\nonumber\\
D_{\rm rot}^\perp&=&\frac{1}{12}n_gR\ell^3 \sqrt{2\pi m \left(k_{\rm B}T\right)^3}\left\{2+12\frac{R^3}{\ell^3}\!+\ac\!\left[\!-\frac{1}{2}+\!\frac{\pi\gamma_{\rm s}}{4}+3\frac{R}{\ell}\!+6\frac{R^2}{\ell^2}+\!\left(\frac{3\pi\gamma_{\rm s}}{2}-6\right)\!\frac{R^3}{\ell^3}\right]\right\}\nonumber\\
D_{\rm rot}^\parallel&=&\frac{1}{2}n_gR^3\ell \sqrt{2\pi m \left(k_{\rm B}T\right)^3}\ac\left(2+\frac{R}{\ell}\right)\nonumber
\end{eqnarray}
} \\ \hline
\rowcolor[gray]{.8}\parbox{0.98\textwidth}{\centering \textbf{Cuboid}} \\ \hline
\parbox{0.18\textwidth}{
\vspace{0.5mm}
\includegraphics[width = 0.13\textwidth]{TableFig3.jpg}
}
\parbox{0.8\textwidth}{\begin{eqnarray}
\mathrm{D}_{\rm cm}&=&D_{\rm cm}^{abc}\bls{n}_a\otimes\bls{n}_a+D_{\rm cm}^{bac}\bls{n}_b\otimes\bls{n}_b+D_{\rm rot}^{cba}\bls{n}_c\otimes\bls{n}_c\nonumber\\
\mathrm{D}_{\rm rot}&=&D_{\rm rot}^{abc}\bls{n}_a\otimes\bls{n}_a+D_{\rm rot}^{bac}\bls{n}_b\otimes\bls{n}_b+D_{\rm rot}^{cba}\bls{n}_c\otimes\bls{n}_c\nonumber\\
&&\nonumber\\
D_{\rm cm}^{abc}&=&\frac{1}{\pi }n_gbc \sqrt{2\pi m\left(k_{\rm B}T\right)^3}\left[4+\ac\left( -2+\frac{ab+ac}{bc}+\frac{\pi\gamma_{\rm s}}{2}\right)\right]\nonumber\\
D_{\rm rot}^{abc}&=&\frac{n_gbc(b^2+c^2)\sqrt{2\pi m(k_{\rm B}T)^3}}{12\pi }\left\{\frac{4a(c^3+b^3)}{bc(b^2+c^2)}+\ac\left[1-\left(2-\frac{\pi}{2}\gamma_{\rm s}\right)\frac{a(c^3+b^3)}{bc(b^2+c^2)}+\frac{3a(b+c)}{b^2+c^2}\right]\right\}\nonumber
\end{eqnarray} } \\ \hline
\end{tabular}
\end{center}
\label{tab:diff}
\end{table*}

\section{Symmetric particles}\label{sec:scc}

In this section we provide the translational and rotational friction and diffusion tensors of spheres, cylinders, and cuboids for specular and diffuse reflection with a constant surface temperature, $T_{\rm s} = {\rm const}$. 
The friction and diffusion tensors \eqref{derivationgammageneral} and \eqref{derivationdiffusiongeneral} are block diagonal  for the considered particle shapes, so that the translational and the rotational thermalization dynamics decouple. 

The corresponding center-of-mass and rotational tensors are presented in Tab.~\ref{tab:fric} and \ref{tab:diff}. All center-of-mass  tensors given in the tables satisfy the relation 
\begin{subequations} \label{flucdiss2}
\begin{equation}
{\rm D}_{\rm cm}(\Omega) = k_{\rm B} T M \Gamma_{\rm cm}(\Omega),
\end{equation}
while all rotational tensors fulfill
\begin{equation}
{\rm D}_{\rm rot}(\Omega) = k_{\rm B} T \Gamma_{\rm rot}(\Omega) \rI(\Omega).
\end{equation}
\end{subequations}
The corresponding tensors  of inertia for spheres, cylinders and cuboids are given by
\begin{subequations}
\begin{align}
 \rI_{\rm sph}(\Omega) = & \frac{2 M R^2}{5} \mathbb{1},
\\
 \rI_{\rm cyl}(\Omega) = & \left( \frac{M \ell^2}{12} + \frac{M R^2}{4} \right ) \left [ \mathbb{1} - {\bf m}(\Omega) \otimes {\bf m}(\Omega) \right ] \nonumber \\ 
  & + \frac{M R^2}{2} {\bf m}(\Omega) \otimes {\bf m}(\Omega),
\\
 \rI_{\rm cub}(\Omega) = & \frac{M}{12} \sum_{k = a,b,c} \left (a^2 + b^2 + c^2 - k^2 \right ) {\bf n}_k(\Omega) \otimes {\bf n}_k(\Omega).
\end{align}
\end{subequations}

For special situations the friction tensors reported in Tab.~\ref{tab:fric} reduce to expressions known in the literature: the translational and rotational friction tensor of a sphere is consistent with Refs.~\cite{epstein1924,halbritter1974,borg2003}, and the center-of-mass friction force on cylinders for $\gamma_{\rm s} = 1$ was calculated in \cite{dahneke1973}, while its frictional torque was determined for $\gamma_{\rm s} = 1$ and for rotations orthogonal to ${\bf m}$ in \cite{li2014}. The friction tensor of cuboids reduces to the one given in Ref.~\cite{cavalleri2009,cavalleri2010} if one takes  $\gamma_{\rm s} = 1$, assumes only diffuse reflection ($\ac=1$), and takes the edge lengths to be equal ($a=b=c$).

Rotational friction and diffusion of spheres occurs only if a finite fraction of the gas atoms is diffusely reflected, since the momentum transfer of specular reflection \eqref{momtransspec} points in the same direction as the surface normal vector. In general, for azimuthally symmetric particles, the angular momentum transfer on rotations around the symmetry axis vanishes unless $\alpha_{\rm c} \neq 0$. This is also evident from the rotational friction tensor for cylinders in Tab.~\ref{tab:fric}, which only describes damping of rotations around ${\bf m}$ if $\ac \neq 0$; see App.~\ref{app:therm}.

\section{Conclusion}\label{sec:conc}

This article provides a comprehensive, unified description of the coupled rotranslational dynamics of rigid objects in thin gases by
deriving a complete classical theory of rotranslational friction, diffusion, and thermalization. While the rotranslational friction matrix is block diagonal for special shapes such as spheres, cylinders an cuboids, this is not the case for arbitrary geometries. Strong rotranslational coupling might give rise to phenomena such as alignment of the particle with its center-of-mass propagation direction or relaxation into nonthermal stationary states. Also the inverse Magnus effect and its rotational analog are manifestations of rotranslational coupling. The field of levitated optomechanics offers a promising platform to observe such effects experimentally.

The strength of the presented formalism lies in its generality and its coordinate-independent formulation in terms of the angular momentum vector. In its general form, the Boltzmann equation makes no assumption on the momentum transferred by a single collision beyond the conservation of linear and angular momentum and the conservation of particles. By slightly modifying the presented theory it will be possible to describe phenomena such as accumulation or outgassing of atoms from the particle surface, an effect that significantly influences the rotation dynamics of comets \cite{whipple1950,gutierrez2003,bodewits2018}. Alternative models of gas-atom surface interaction, potentially motivated by ab initio calculations, can be easily included by adapting the momentum transfer function.

Beyond its applications in classical physics, the here derived translational and rotational friction and diffusion tensors can be used in phenomenological models of friction and diffusion of quantum rotors \cite{caldeira1983,papendell2017,stickler2017} and thereby contribute to decoherence experiments and noninterferometric tests of objective collapse models \cite{goldwater2016,li2016,schrinski2017,helou2017}.

\appendix

\section{Phase space formulation} \label{app:tdf}

The canonically conjugate momenta to the Euler angles $\Omega = (\alpha,\beta,\gamma)$ in the $z$-$y'$-$z''$ convention are denoted by  $p_\Omega = (p_\alpha, p_\beta,p_\gamma)$. They are related to the angular momentum vector ${\bf J}$ via
\begin{subequations}\label{appendixtrafoeulersmomenta}
\begin{equation}
p_\alpha=\bls{J}\cdot\bls{e}_z=\Phi_\alpha(\Omega,\bls{J}),
\end{equation}
\begin{equation}
p_\beta=\bls{J}\cdot\bls{e}_\xi(\Omega)=\Phi_\beta(\Omega,\bls{J}),
\end{equation}
\begin{equation}
p_\gamma=\bls{J}\cdot{\bf m}(\Omega)=\Phi_\gamma(\Omega,\bls{J}),
\end{equation}
\end{subequations}
where ${\bf e}_\xi(\Omega) = - \sin \alpha {\bf e}_x + \cos \alpha {\bf e}_y$ and ${\bf m}(\Omega) = \cos \alpha \sin \beta {\bf e}_x + \sin \alpha \sin \beta {\bf e}_y + \cos \beta {\bf e}_z$, and we denoted the transformation functions by $\Phi_\Omega(\Omega,{\bf J})$.

The phase space density $h_t(\Omega,p_\Omega)$ is related to the probability density $f_t(\Omega,{\bf J})$ used in the main text by the (norm-preserving) transformation
\begin{align}
h_t(\Omega,p_\Omega)=&\int d^3\!J\,f_t(\Omega,\bls{J})\,\delta\!\left[p_\Omega-\Phi_{\Omega}(\Omega,\bls{J})\right].
\end{align}
Similarly, the inverse relation is
\begin{equation}
f_t(\Omega,\bls{J})= \int d^3\!p_\Omega\,h_t(\Omega,p_\Omega)\,\delta\!\left[\bls{J}-\boldsymbol{\Psi}_{ J}(\Omega,p_\Omega)\right],
\end{equation}
where $ \boldsymbol{\Psi}_{ J}(\Omega,p_\Omega)$ is the inversion of Eqs.~\eqref{appendixtrafoeulersmomenta} so that $\Phi_\Omega[\Omega,\boldsymbol{\Psi}_{ J}(\Omega,p_\Omega)] = p_\Omega$. The integrals can be carried out, yielding
\begin{equation}
f_t(\Omega,\bls{J})=\sin\beta\, h_t\left[\Omega,\Phi_{\Omega}\!\left(\Omega,\bls{J}\right)\right ],
\end{equation}
where $\sqrt{g(\Omega)} = \sin \beta$ is the Jacobian determinant of the transformation. Consequently, the normalization of $f_t$ and $h_t$ reads
\begin{subequations}
\begin{align}
 \int d^3 \Omega \int d^3 p_\Omega h_t(\Omega,p_\Omega) = & 1, \\
 \int d^3 \Omega \int d^3 J f_t(\Omega,{\bf J}) = & 1,
\end{align}
\end{subequations}
with $d^3 \Omega = d \alpha d\beta d\gamma$.

In the absence of gas collisions the phase space distribution satisfies the Liouville equation
\begin{equation}
\partial_t h_t+\{h_t,H\}=0, \label{appentropychangeliouville}
\end{equation}
with the rigid rotor Hamilton function in terms of phase space variables
\begin{align}
H=&\frac{\bls{P}^2}{2M}\!+\!\frac{\left[\left(p_\alpha\!-\!p_\gamma \cos\beta\right)\cos\gamma-p_\beta\sin\beta\sin\gamma\right]^2}{2I_1\sin^2\beta} \nonumber\\
&+\frac{\left[\left(p_\alpha-p_\gamma \cos\beta\right)\sin\gamma+p_\beta\sin\beta\cos\gamma\right]^2}{2I_2\sin^2\beta}\nonumber\\
&+\frac{p_\gamma^2}{2I_3}+V(\bls{R},\Omega).
\end{align}
The free time evolution for $f_t(\Omega,{\bf J})$ is rather complicated in general and can be obtained by transformation of the Poisson bracket.

\section{Thermalization for specular and diffuse reflection}\label{app:therm}

The Fokker-Planck equation \eqref{fokkerplanckgleichung} for specular and diffuse reflection with $T_{\rm s} = T$ predicts that an arbitrary initial state approaches the stationary solution \eqref{feq}. This can be demonstrated \cite{gardiner1985,risken1996} by considering the relative entropy
\begin{equation}
 S(t) = - \int d^3\!Rd^3\kern-0.05em\Omega d^3\!Pd^3\!p_\Omega\, h_t\ln\left(\frac{h_t}{h_{\rm eq}}\right)
\end{equation}
in terms of the phase space distribution discussed in App.~\ref{app:tdf}.
This entropy is always negative \cite{risken1996} and it is conserved under the free dynamics since
\begin{equation}
\int d^3\!Rd^3\kern-0.05em\Omega d^3\!Pd^3\!p_\Omega\,\{h_t,H\}\,{\rm ln}\left(\frac{h_t}{h_{\rm eq}}\right)=0.
\end{equation}

In order to calculate the effect of gas collisions on the entropy, we note that $S(t)$ can be written as
\begin{equation}
S(t)=-\int d^3\!Rd^3\kern-0.05em\Omega d^3\!Pd^3\!J\, f_t\ln\left(\frac{f_t}{f_{\rm eq}}\right),\label{appendixentropyviaf}
\end{equation}
see App.~\ref{app:tdf}, 
leading to
\begin{equation}
\frac{d}{dt}S(t)= -\int d^3\!Rd^3\kern-0.05em\Omega d^3\!Pd^3\!J\, \left({\partial}_t^{\rm coll}f_t \right )\ln\left(\frac{f_t}{f_{\rm eq}}\right).
\end{equation}
This can be further evaluated by inserting the Fokker-Planck equation \eqref{fokkerplanckgleichung} with $T_{\rm s}=T$ and integrating by parts. Using that $\partial_t^{\rm coll}f_{\rm eq}$=$0$, one finds
\begin{align}
\frac{d}{dt}S(t)=&\int d^3\!Rd^3\kern-0.05em\Omega d^3\!Pd^3\!J\,f_t\left[\left(\begin{array}{c} \!\!\nabla_{\bls{P}}\!\! \\ \!\!\nabla_{\bls{J}}\!\!\end{array}\right)\ln\left(\!\frac{f_t}{f_{\rm eq}}\!\right)\right]\cdot \nonumber\\
&\cdot \mathsf{D}(\Omega)\left [\left(\begin{array}{c} \!\!\nabla_{\bls{P}}\! \!\\ \!\!\nabla_{\bls{J}}\!\!\end{array}\right)\ln\left(\!\frac{f_t}{f_{\rm eq}}\!\right)\right].
\end{align}
This is never negative since the diffusion matrix ${\sf D}(\Omega)$ is positive semidefinite according to its definition \eqref{derivationdiffusiongeneral}. 

Below we will demonstrate that ${\sf D}(\Omega)$ is positive definite for $\alpha_{\rm c} \neq 0$, implying that the entropy increases monotonically until
\begin{equation}
\left(\begin{array}{c} \!\!\nabla_{\bls{P}}\! \!\\ \!\!\nabla_{\bls{J}}\!\!\end{array}\right)\ln\left(\!\frac{f_t}{f_{\rm eq}}\!\right)=0\label{thermalizationequationtoeqfunct}
\end{equation}
for all but isolated points. Thus, the entropy is only stationary if $f_t \equiv f_{\rm eq}$. However, stationarity of $S(t)$ is necessary but not sufficient for the stationarity of $f_t$. In the above argument we exploited that $\partial_t^{\rm coll} f_{\rm eq} = 0$, which is true for all distributions of the form \eqref{statsol}. The additional requirement that $f_{\rm eq}$ is invariant under the free dynamics \eqref{appentropychangeliouville} determines its prefactor and finally leads to the unique thermal state \eqref{feq}. This is illustrated for the linear rigid rotor in Fig.~\ref{picsimphasenraum}.

We now demonstrate that the diffusion matrix \eqref{derivationdiffusionspecular} is indeed positive definite for $\ac \neq 0$. Equation \eqref{derivationdiffusionspecular} contains two terms, both integrated over the particle surface and multiplied by positive prefactors. The first is proportional to
\begin{equation}\label{quadform} 
\left(\begin{array}{rr} \mathbb{1}\,\, & -\mathsf{r} \\
     \mathsf{r}\,\, & -\mathsf{r}^2\end{array}\right)=
     \left(\begin{array}{rr} \mathbb{1}\,\, & \mathbb{0} \\
     \mathsf{r}\,\, & \mathbb{0}\end{array}\right)
     \left(\begin{array}{rr} \mathbb{1}\,\, & -\mathsf{r} \\
     \mathbb{0}\,\, & \mathbb{0}\end{array}\right)=\mathsf{G}\mathsf{G}^{\rm T},
\end{equation}
where its Choleksy decomposition in terms of ${\sf G}$ implies that it is positive semidefinite. By solving the quadratic equation
\begin{equation} \label{quadform2}
{\sf v} \cdot{\sf G}^{\rm T} {\sf G} {\sf v}=0,
\end{equation}
for ${\bf r}$ one finds that for a given vector ${\sf v}$ the solutions lie along a straight line, but never on a closed surface. The quadratic form in \eqref{quadform2} is thus positive for every vector $\bls{r}$ off this line and, hence, integration over a closed surface always yields a positive definite matrix,
\begin{equation}\label{pd}
\int_{\partial V}dA\left(\begin{array}{rr} \mathbb{1}\,\,&  -\mathsf{r}\\
     \mathsf{r}\,\, & -{\sf r}^2\end{array}\right) > 0.
\end{equation}

The second term in Eq.~\eqref{derivationdiffusionspecular} is positive semidefinite since
\begin{equation}\label{quadformxxx}
{\sf v} \cdot \left(\begin{array}{c} \!\!\boldsymbol{n}\!\! \\ \!\!\boldsymbol{r}\times\boldsymbol{n}\!\!\! \end{array}\right)\otimes\left(\begin{array}{c} \!\!\boldsymbol{n}\!\! \\ \!\!\boldsymbol{r}\times\boldsymbol{n}\!\!\! \end{array}\right) {\sf v} \geq 0.
\end{equation}
and remains so even after integrating over a closed surface. 
Eqs.~\eqref{pd} and \eqref{derivationdiffusionspecular} thus show that the diffusion matrix is positive definite if $\ac\neq0$, i.e., if a fraction of the gas atoms is reflected diffusely. Any initial state therefore evolves towards \eqref{feq}.

However, in the case of pure specular reflection thermalization does not necessarily take place. This can be illustrated by investigating the diffusion matrix of spheres ($\ac=0$)
\begin{equation}
{\sf D}_{\rm sph}(\Omega)\propto\left(\begin{array}{rr} \mathbb{1}\,\,&  \mathbb{0}\\
     \mathbb{0}\,\, & \mathbb{0}\end{array}\right),
\end{equation}
whose inverse does not exist (rendering Eq.~\eqref{statsol} invalid). Spheres therefore thermalize only in their center-of-mass and not in their rotational degrees of freedom if $\ac = 0$. The same holds for azimuthally symmetric particles and their rotations around the symmetry axis, as also reflected in the friction and diffusion tensors for cylinders in Tabs.~\ref{tab:fric} and \ref{tab:diff}.

\end{document}